\documentclass[12pt]{article}
\usepackage{amsmath}
\usepackage{tikz}
\usetikzlibrary{positioning}
\usepackage{graphicx,psfrag,epsf}
\usepackage{enumerate}
\usepackage{natbib}
\usepackage{url} 
\usepackage{xcolor}
\usepackage{algorithm}
\usepackage{algpseudocode}
\usepackage{booktabs}
\usepackage{amsthm}
\usepackage{changes}

\newtheorem{theorem}[]{Theorem}

\newcommand\reps{500}
\newcommand\ntimes{100}
\newcommand\truereps{2000}
\newcommand\datareps{5}
\long\def\ignore#1{}

\DeclareMathOperator{\inprob}{\overset{p}{\to}}
\newcommand{\blind}{1}

\usepackage{shortcuts}

\begin{document}

\def\spacingset#1{\renewcommand{\baselinestretch}%
{#1}\small\normalsize} \spacingset{1}


\if1\blind
{
  \title{\bf A Fast Bootstrap Algorithm for Causal Inference with Large Data}
  \author{Matthew Kosko\\
    Department of Statistics, George Washington University,\\
    Washington, DC, 20052\\
    Lin Wang \\
    Department of Statistics, Purdue University,\\
    West Lafayette, IN 47907, \\
    and \\
    Michele Santacatterina \\
    Department of Population Health, New York University, \\ 
    New York, NY, 10016}
  \maketitle
} \fi

\if0\blind
{
  \bigskip
  \bigskip
  \bigskip
  \begin{center}
    {\LARGE\bf A Fast Bootstrap Algorithm for Causal Inference with Large Data}
\end{center}
  \medskip
} \fi

\bigskip
\begin{abstract}

Estimating causal effects from large experimental and observational data has become increasingly prevalent in both industry and research. The bootstrap is an intuitive and powerful technique used to construct standard errors and confidence intervals of estimators. Its application however can be prohibitively demanding in settings involving large data. In addition, modern causal inference estimators based on machine learning and optimization techniques exacerbate the computational burden of the bootstrap. The bag of little bootstraps has been proposed in non-causal settings for large data but has not yet been applied to evaluate the properties of estimators of causal effects. In this paper, we introduce a new bootstrap algorithm called causal bag of little bootstraps for causal inference with large data. The new algorithm significantly improves the computational efficiency of the traditional bootstrap while providing consistent estimates and desirable confidence interval coverage. We describe its properties, provide practical considerations, and evaluate the performance of the proposed algorithm in terms of bias, coverage of the true 95\% confidence intervals, and computational time in a simulation study.  We apply it in the evaluation of the effect of hormone therapy on the average time to coronary heart disease using a large observational data set from the Women's Health Initiative.

\end{abstract}


\noindent%
\vfill

\newpage
\spacingset{1.45} 
\section{Introduction}
\label{sec:intro}

In recent years, large experimental and observational data sets aimed at inferring causal relationships have become increasingly prevalent in both industry \citep{kohavi2009online,tang2010overlapping,peysakhovich2016combining,kohavi2013online} and medical research  \citep{dagan2021bnt162b2,mohammadi2021marginal,sud2020association}. For instance, \cite{tang2010overlapping} evaluated user-visible changes and machine-learning algorithms using large A/B tests, while \cite{dagan2021bnt162b2} evaluated the effectiveness of an mRNA COVID-19 vaccine on a population of more than one million participants. 

There are a variety of methods to estimate causal effects. Techniques based on the propensity score, such as the inverse probability weighting (IPW) estimator \citep{lunceford2004stratification} and covariate balancing propensity score (CBPS) \citep{imai2014covariate} are commonly deployed. IPW weights are constructed as the inverse of the propensity score \citep{rosenbaum1983central} which can be estimated using both parametric and nonparametric machine learning techniques such as support vector machines (SVM) \citep{westreich2010propensity}. Alternatively, CBPS finds the logistic model that balances covariates via the generalized method of moments \citep{imai2014covariate}. 

The bootstrap \citep{eforn1979bootstrap,efron1994introduction} 
has been used to obtain estimates of precision for statistical estimators including standard errors and confidence intervals. 
While the bootstrap's popularity has grown with increased access to modern computing power, its application can still be prohibitively demanding in settings involving large data sets.  

Methods have been proposed to improve the computational efficiency of the bootstrap. \cite{politis1999subsampling} and \cite{bickel2012resampling} proposed sub-sampling and the closely related $m$ out of $n$ bootstrap, respectively. These methods obtain bootstrap estimates on smaller subsamples, thus improving on the standard bootstrap's computational limitations.  These procedures, however, have their own disadvantages \citep{samworth2003note,kleiner2014scalable}. In particular, their finite sample behavior is poor relative to the bootstrap and is sensitive to the choice of subsample size. 
To overcome these challenges, \cite{kleiner2014scalable} proposed the bag of little bootstraps (BLB), which, instead of applying an estimator to a smaller subsample, deploys the bootstrap on multiple subsets or ``bags'' of the data. Finally, BLB draws bootstrap samples equal to the size of the full dataset. While BLB improves the scalability of the bootstrap, it has not yet been extended to obtain estimates
of precision for 
estimators of \textit{causal} effects. 

In this paper, we introduce a new bootstrap algorithm called causal bag of little bootstraps (causal BLB) for causal inference with large data. Our work was motivated by the growing number of large data sets inferring causal relationships and by the use of machine learning and covariate balancing techniques to estimate causal effects. 
In the following section, we provide an overview of existing work related to the bootstrap as applied to causal inference and our contributions to this literature. We introduce our proposed method and discuss its properties in section \ref{sec:causalBLB}. We provide practical considerations in section \ref{sec:pract_guidelines} and evaluate the method performance with respect to computational time, bias, and coverage of the 95\% confidence intervals in section \ref{sec:simulate}. We apply the proposed method in the evaluation of the effect of hormone therapy on time to coronary heart disease using a large observational dataset from the Women's Health Initiative study. We provide conclusions in section \ref{sec:conc}.

\subsection{Related work}

Although there is a wide literature on applying the bootstrap to a variety of statistical problems \cite[among many others]{wu2022low,zhu2020novel}, the literature on the bootstrap in causal inference is relatively small, especially for bootstrapping for causal inference with large data. The usual method for employing the bootstrap in causal problems involves taking a sample with replacement from the data and ``re-designing'' each sample to ensure covariate balance \citep{zhang2021designed,dagan2021bnt162b2}. 

There is a more significant literature on the use of bootstrap for matching estimators. \cite{abadie2008failure} and \cite{abadie2022robust} both look at the bootstrap in the context of matching estimators. \cite{abadie2008failure} show that the standard bootstrap does not produce valid inference for matching estimators and requires modification. Along this line,  \citet{abadie2022robust} examine how to construct valid standard errors for regression coefficients, including treatment coefficients, after matching. The authors develop a block version of the nonparametric bootstrap that resamples matched sets rather than individual observations; they show that this procedure produces valid inference. \cite{otsu2017bootstrap} also overcomes the problem of the standard bootstrap for matching estimators found by \citet{abadie2008failure}; the authors construct a weighted bootstrap procedure that does not recompute the number of times an observation is used in calculating the bootstrap estimator; rather, it is resampled as part of the observations. \cite{adusumilli2018bootstrap} proposes a modified bootstrap that relies on the concept of potential errors. \cite{zhao2019sensitivity} applied the bootstrap to sensitivity analysis, deriving confidence intervals for sensitivity analysis. 

Some work has been done to evaluate the performance of bootstrap in causal inference. \cite{austin2016variance} evaluated the performance of the bootstrap in estimating the variance of marginal hazard ratios using a weighted-Cox model in a simulation study. In another simulation study, \cite{austin2014use} showed the performance of the bootstrap when using propensity-score matching without replacement for estimating average treatment effects.



Recent rigorous work in bootstrapping for causal inference has been done by 
\cite{imbens2021causal}. They introduce a causal bootstrap algorithm for causal inference with both observational and randomized data and establish favorable large-sample properties. Particularly, they show that the bootstrap confidence intervals are asymptotically conservative. The procedure from \cite{imbens2021causal} first imputes the potential outcomes for the treated and control groups using a copula function that maps the marginal density functions to the joint density. It then simulates both the sampling and randomization distributions by drawing from the original data and simulating a new treatment vector. Using a particular copula, the authors show this procedure produces conservative inference. For observational data, a weighted empirical cumulative density function 
that incorporates a balancing score is used in the imputation step to ensure that treatment is independent of the potential outcomes. Despite the large sample appeal of this method, the imputation step required becomes computationally intensive with large datasets. 

\subsection{Our contribution}

Our contribution to this field of literature is to provide a novel bootstrap algorithm that improves the computational time of the traditional bootstrap while consistently estimating causal effects from large data with desirable confidence interval coverage.  In contrast to \cite{imbens2021causal}, our proposed method does not require imputation and instead relies on drawing weighted bootstrap samples. 
By doing so, we significantly increase computational efficiency while maintaining an intuitive interpretation of the method as repeated draws of a weighted estimator. \if1\blind
{In addition, we provide the \texttt{R} code for the algorithm at \url{https://github.com/mdk31/causalbootstrap}.
} \fi 





\section{Causal Bag of Little Bootstrap}
\label{sec:causalBLB}

We consider an experimental or observational study consisting of $n$ units, drawn independently and identically distributed (iid). 
Using the potential outcome framework \citep{imbens2015causal}, for each unit $i=1,\dots,n$, we let $Y_i(w) \in \mathbf{R}$ be the potential outcome of treatment $w \in \lbrace 0,1\rbrace$. We let $X_i \in \mathcal{X}$ be the observed confounders.  We set $W_i$ the indicator of being treated with treatment $w$. In this paper, our focus is on estimating the average treatment effect (ATE), given by

\[
\tau \equiv E(Y(1) - Y(0)),
\]

\noindent
which is identifiable by assuming consistency, non-interference, and ignorable treatment assignment \citep{imbens2015causal,hernan2020causal, rosenbaum1983central}. 


In this paper, we are interested in using the bootstrap to compute standard errors and confidence intervals. 
As previously mentioned, when $n$ increases the bootstrap becomes prohibitively demanding. We now describe the causal bag of little bootstraps (causal BLB) a fast bootstrap algorithm that extends the original algorithm introduced in \citet{kleiner2014scalable} to causal inference. The causal BLB algorithm starts by drawing $s$ subsets of size $b$ from the original data. It then obtains a set of weights by using parametric, machine learning, or covariate balancing techniques for each subset. The algorithm then takes $r$ bootstrap resamples of size $n$ within each subset using its weights and, for the $j$th bootstrap resample drawn from the $k$th subset, it estimates $\tau$ using the following weighted estimator:

\[
 \hat{\tau}_{j,k} =\sum_{i=1}^{b}\left(\dfrac{-1}{n_0}\right)^{W_{i,k} - 1}\left(\dfrac{1}{n_1}\right)^{W_{i,k}}\Tilde{M}_{i,j,k}Y_{i,k},
 \]
\noindent
where 
$n_0$ and $n_1$ are the number of control and treated units in the full dataset, $\Tilde{M}_{j,k}$ is a vector made by concatenating two vectors $M^1$ and $M^0$, the treatment and control multinomial draws respectively within the $j$-th bootstrap replicate with probabilities equal to the normalized propensity-score-based weights: $$(\hat{w}^0(X_i), \hat{w}^1(X_i)) = \left(\frac{1/(1-\hat{\pi}(X_i))}{\sum_{i=1}^n (1-W_i)/(1-\hat{\pi}(X_i))}, \frac{1/\hat{\pi}(X_i)}{\sum_{i=1}^n W_i/\hat{\pi}(X_i)} \right),$$ where $\hat{\pi}(X_i)$ is the estimated propensity score. Here, the bootstrap oversamples the data in both the treatment and the control group so the sum of the elements of $M^1$ and $M^0$ is, respectively, $n_1$ and $n_0$. Because an observation can appear multiple times in any bootstrap resample, we recast the bootstrap as integer multiplication of each observation in the subset, representing how many times the obserfvations appears in the bootstrap resample and where the integer multinomial distributed, as in \cite{praestgaard1993exchangeably}. This ensures that the size of the total bootstrap resample is $n = n_0 + n_1$.

\begin{figure}
    \centering
    \includegraphics[scale = 0.45]{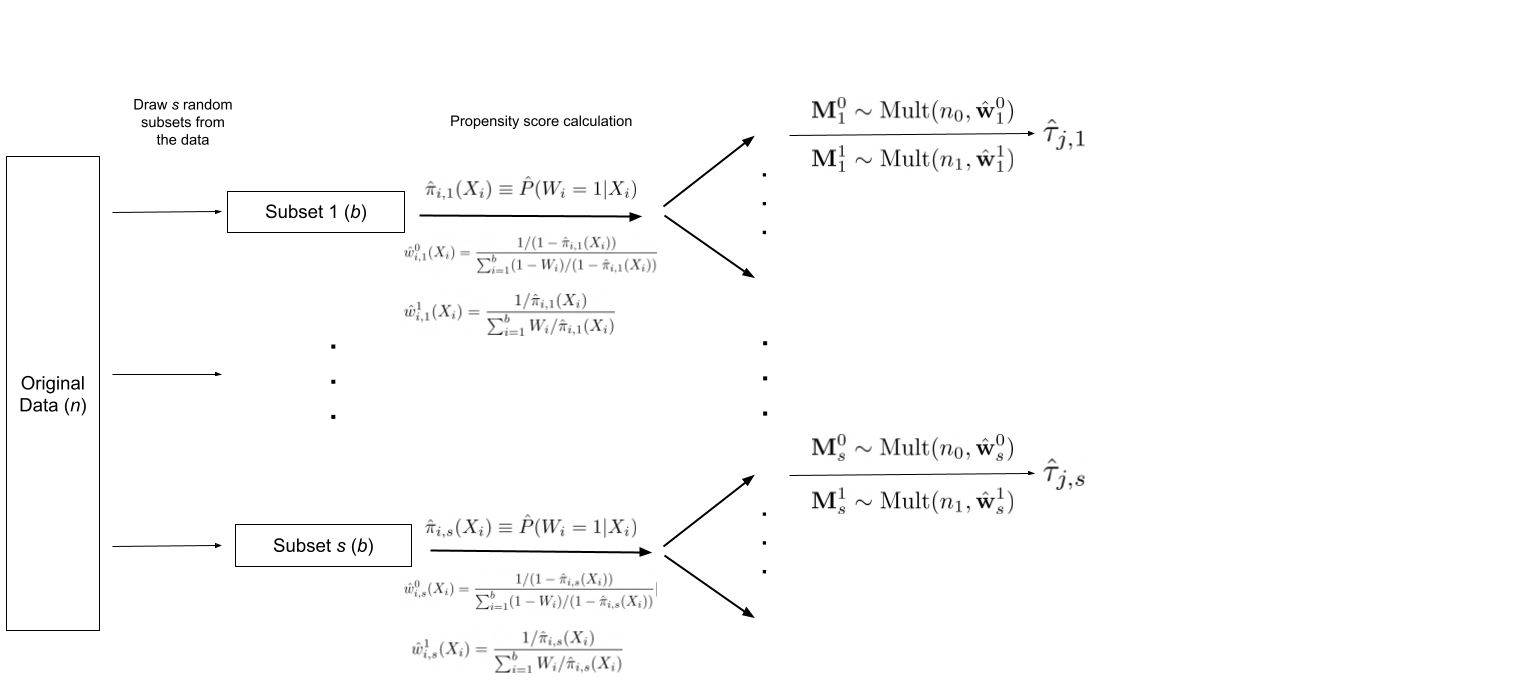}
    \caption{A graphical representation of the causal BLB algorithm.} 
    \label{fig:causalblbfig}
\end{figure}

Finally, the causal BLB algorithm  averages $\hat{\tau}_{j,k}$ across resamples and across subsets to obtain an overall bootstrap estimate given by:
\begin{align*}
   \hat{\tau} &= \dfrac{1}{s}\sum_{k=1}^s\dfrac{1}{r}\sum_{j=1}^r\hat{\tau}_{j,k} \\
   &=\dfrac{1}{s}\sum_{k=1}^s\dfrac{1}{r}\sum_{j=1}^r\sum_{i=1}^{b}\left(\dfrac{-1}{n_0}\right)^{W_{i,k} - 1}\left(\dfrac{1}{n_1}\right)^{W_{i,k}}\Tilde{M}_{i,j,k}Y_{i,k}. 
\end{align*}
\noindent
By doing so, the causal BLB algorithm inherits the conventional way of employing the bootstrap in causal problems, \textit{i.e.}, taking a weighted sample from the data thus ensuring covariate balance as in \cite{zhang2021designed} and \cite{dagan2021bnt162b2}. The causal BLB is summarized in Algorithm \ref{alg:cblb} and Figure \ref{fig:causalblbfig}. 




\begin{algorithm}
\begin{algorithmic}
\Require Number of subsets $s$, their size $b$, and number of bootstrap replicates, $r$
\State Calculate $n_1 = \sum_{i=1} W_i$ and $n_0 = n - n_1$
\For{$k \gets 1$ to $s$}
    \State Sample a set of indices $I_k = \{i_1, \ldots, i_b\}$ from $\mathcal{I} = \{1, 2, \ldots, n\}$ without replacement
    \State For the subset data $\mathbf{O}_k = (W_{i_1}, \ldots, W_{i_b},X_{i_1}, \ldots, X_{i_b})$,
    construct a model  of the propensity score, $\hat{\pi}_{i,k}(X_i)$ and balance the data
    \State Reorder the observations such that the first $b_k^0 = \sum_{l=1}^b (1-W_{i_l})$ are controls and the last $b - b_k^0$ are treated 
    \State Construct normalized inverse propensity weights for the subset data 
    \State $(\hat{w}_{i,k}^0(X_i), \hat{w}_{i,k}^1(X_i)) = \left(\frac{1/(1-\hat{\pi}_{i,k}(X_i))}{\sum_{i=1}^b (1-W_i)/(1-\hat{\pi}_{i,k}(X_i))}, \frac{1/\hat{\pi}_{i,k}(X_i)}{\sum_{i=1}^b W_i/\hat{\pi}_{i,k}(X_i)} \right)$
    \For{$j \gets 1$ to $r$}
        \State Sample $\left(M_{1, j, k}^1, \ldots, M_{b_k^1, j, k}^1\right) \sim \textrm{Multinomial}(n_1, \hat{\mathbf{w}}_k^1)$
        \State Sample $\left(M_{1, j, k}^0, \ldots, M_{b_k^0, j, k}^0\right) \sim \textrm{Multinomial}(n_0, \hat{\mathbf{w}}_k^0)$
        \State Compute $\Tilde{M}_{j,k} = (M_{1, j, k}^0, \ldots, M_{b_k^0, j, k}^0, M_{1, j, k}^1, \ldots, M_{b_k^1, j, k}^1)$
        \State Calculate $\hat{\tau}_{j,k} =\sum_{i=1}^{b}\left(\dfrac{-1}{n_0}\right)^{W_{i,k} - 1}\left(\dfrac{1}{n_1}\right)^{W_{i,k}}\Tilde{M}_{i,j,k}Y_{i,k}$
    \EndFor
    \State $\hat{\tau}_{k} \gets \frac{1}{r}\sum_{j=1}^r \hat{\tau}_{j,k}$
    \State $\textrm{se}(\hat{\tau}_{k}) \gets \sqrt{\frac{1}{r-1}\sum_{j=1}^r \left(\hat{\tau}_{j,k} - \hat{\tau}_k\right)^2}$
    \State $\textrm{CI}(\tau)_k \gets (\hat{\tau}_{k}^{0.025}, \hat{\tau}_{k}^{0.975})$ (2.5\% and 97.5\% percentiles of the bootstrap draws)
\EndFor
\State $\hat{\tau} \gets \frac{1}{s}\sum_{k=1}^s \hat{\tau}_k$ 
\State $\textrm{se}(\hat{\tau}) \gets \frac{1}{s}\sum_{k=1}^s \textrm{se}(\hat{\tau}_k)$
\State $\textrm{CI}(\tau) \gets \left(\frac{1}{s}\sum_{k=1}^s\hat{\tau}_{k}^{0.025}, \frac{1}{s}\sum_{k=1}^s\hat{\tau}_{k}^{0.975}\right)$
\end{algorithmic}
\caption{Causal BLB}\label{alg:cblb}
\end{algorithm}

\subsection{Properties}
\label{sec:properties}

In this section, we discuss some properties of causal BLB. Specifically, we start by showing that causal BLB provides consistent estimates of ATE, $\tau$. We then discuss connections between our proposed causal BLB and the causal bootstrap algorithm proposed by \cite{imbens2021causal}. We finally discuss time complexity improvements by comparing the causal BLB to the traditional bootstrap.

\begin{theorem}[Consistency]
Under (causal) consistency, non-interference, ignorable treatment assignment, and correct specification of the propensity score model, $\hat{\tau} \inprob \tau$. 
\end{theorem}

\noindent
The proof is provided in section  \ref{sec:consistent} of the appendix. In summary, we show that the causal BLB estimator behaves similarly to the variance-stabilized IPW estimator; in particular, we show that it converges in probability to the IPW estimator as long as the propensity score model is correct. 

\paragraph{Connections with the causal bootstrap method of \cite{imbens2021causal}}

\cite{imbens2021causal} develop a causal bootstrap by taking the uncertainty as arising from the stochastic nature of the assignment instead of the sampling uncertainty. They impute the joint distribution of the potential outcomes for the control and treated groups using the isotone copula
$
\hat{F}_{01}^{iso}=C^{iso}(\hat{F}_0,\hat{F}_1),
$
where $\hat{F}_0$ and $\hat{F}_1$ are the marginal empirical cumulative distribution functions (CDFs) for the control and treated groups respectively. Specifically, for randomized experiments, they simply impute the missing counterfactuals according to
\begin{align*}
\tilde{Y}_i(0) &:= \begin{cases} Y_i & \textrm{if } W_i = 0,\\ \hat{F}_0^{-1}(\hat{F}_1(Y_i)) & \textrm{otherwise,} \end{cases} \\
\tilde{Y}_i(1) &:= \begin{cases} Y_i & \textrm{if } W_i = 1,\\ \hat{F}_1^{-1}(\hat{F}_0(Y_i)) & \textrm{otherwise.} \end{cases} 
\end{align*}
Then they simulate the randomization distribution by repeatedly drawing $n$ units out of the imputed empirical population. They also show that an estimator for the distribution of $\hat{\tau}_{ATE}$ which assumes the isotone coupling is asymptotically conservative at any order of approximation.

Such a causal bootstrap framework may work well for samples of small or moderate size but is computationally prohibitive for big data due to the imputation, especially when propensity score estimation and kernel imputation are needed for observational designs. In fact, it is known that the isotone copula is attained when $Y_i(1)=T(Y_i(0))$, where $T$ is a strictly increasing transformation, (see, for example, \cite{schmidt2007coping}). We now show that under the assumption of $Y_i(1)=T(Y_i(0))$, the sampling uncertainty is equivalent to the assignment uncertainty so that our sampling-based bootstrap framework is equivalent to the causal bootstrap of \cite{imbens2021causal}. We only demonstrate this for experimental trials, and the case for observational designs can be shown similarly just with tedious notations. 
Note that $\hat{F}_1(Y_i(1))=\hat{F}_0(T^{-1}(Y_i(1)))=\hat{F}_0(Y_i(0))$, and the imputed CDF of $\tilde{Y}_i(0)$ is given by
\begin{align*}
Pr(\tilde{Y}_i(0)\leq y)
&=\frac{n_0}{n}Pr(Y_i\leq y|W_i=0)+\frac{n_1}{n}Pr(\hat{F}_0^{-1}(\hat{F}_1(Y_i))\leq y|W_i=1)\nonumber\\
&=\frac{n_0}{n}Pr(Y_i\leq y|W_i=0)+\frac{n_1}{n}Pr(\hat{F}_1(Y_i)\leq \hat{F}_0(y)|W_i=1)\nonumber\\
&=\frac{n_0}{n}Pr(Y_i\leq y|W_i=0)+\frac{n_1}{n}Pr(\hat{F}_0(Y_i)\leq \hat{F}_0(y)|W_i=0)\\
&=\hat{F}_0(y), \nonumber
\end{align*}
which is the empirical CDF of the observed $Y_i(0)$. 
We can show similarly that the imputed CDF of $\tilde{Y}_i(1)$ equals the empirical CDF of the observed $Y_i(1)$. Therefore, drawing from the imputed population is equivalent to drawing from the sample.


\paragraph{Improved time complexity}

Table \ref{tab:complexity} shows the time complexity for methods commonly used to obtain weights for weighted estimators in causal inference using causal BLB. As in \cite{kleiner2014scalable}, we write the subset size $b$ using a parameter $\gamma$ so $b = n^\gamma$ (appropriately rounded so $b$ is an integer). For example, when $\gamma = 0.7$ and $n = 10000$, $b = 631$.

We see that, for both CBPS and SVM, the time complexities suggest a distinct time advantage in using the causal BLB method while there is little to be gained using standard logistic regression. This is indeed what we see when we examine the simulation time elapsed in Figure \ref{fig:methodtime}. Note that when the size of the subset is too small relative to the data size, iterative algorithms can run longer and data splitting can aggravate timing issues. See our discussion of this and CBPS in section \ref{sec:whiresults}.

\begin{table}
    \centering
    \begin{tabular}{c c c c}
    \toprule
      Method & Solver & Traditional TC & Causal BLB TC \\ 
      \midrule
        Logistic regression & IRLS & $O(n)$ & $O(sn^\gamma)$ \\
        CBPS & BFGS & $O(n^2)$ & $O(sn^{2\gamma})$\\
        SVM & SMO & $O(n^3)$ & $O(sn^{3\gamma})$ \\
        \bottomrule
    \end{tabular}
    \caption{Time complexity (TC) of algorithms used in causal inference.}
    \label{tab:complexity}
\end{table}

\paragraph{Data-structure agnostic} Causal BLB is data-structure agnostic in the sense that it can be applied to both randomized and observational data. For instance, one can use causal BLB to estimate ATE from large experiments by either estimating propensities as the marginal or conditional probabilities of treatment assignment \citep{lunceford2004stratification}.

\section{Practical Considerations}
\label{sec:pract_guidelines}

The causal BLB algorithm depends on the method for obtaining propensity scores and its hyperparameters, the number of subsets $s$ and the size of the subsets $b$, the number of bootstrap samples, and the type of confidence interval. In this section, we provide some practical guidelines on their choice. For choice of hyperparameters, our recommendations are based on relative error simulations described in section \ref{sec:hyperparam} of the supplementary material.

\paragraph{Method for estimating propensity scores.}

Although logistic regression is the most commonly used method for estimating propensity scores in IPW estimators, there are a variety of algorithms that can be used to obtain them; these include both nonparametric and machine learning methods \citep{lee2010improving}. The advantage of the causal BLB is that it does not require the use of any particular model however, as discussed in section \ref{sec:properties}, certain algorithms have distinct time advantages due to their complexity.

In the simulation presented in section \ref{sec:simulate}, we used three different methods to estimate the propensity score weights, namely logistic regression, a support vector machine (SVM) with a linear kernel and cost parameter equal to 0.01, and CBPS.

All three methods work reasonably well when the number of subsets is small relative to the data. However, Figure \ref{fig:methodtime} shows that the size of each subset can radically effect computation time. In particular, using CBPS and SVM with causal BLB using relatively small $b$ and large $s$ substantially reduces computation time. Additionally, hyperparameter tuning that may be computationally prohibitive on the scale of the full dataset should be run on individual subsets instead.




\paragraph{Size of the subsets $b$.} Choosing the size of each subset and the number of subsets are intimately connected and depend on one another. In general, using fewer observations per subset requires more subsets to obtain adequate results. In the original BLB paper, \cite[p. 20]{kleiner2014scalable} recommend subset sizes $b$ on the basis of a parameter $\gamma$, where $b = n^{\gamma}$. Based on the many simulations they perform, the authors find that $\gamma = 0.7$ is a ``a reasonable and effective choice'' in many situations \citep[p. 20]{kleiner2014scalable}. In our own simulations (see relative error trajectories of Figure \ref{fig:svmerr} ad Figure \ref{fig:logiterr} of the supplementary material) for the causal BLB, we find that the appropriate $\gamma$ changes by estimating method. For data-intensive machine-learning methods like SVM, adding subsets is not sufficient for the causal BLB algorithm to converge to a low relative error. These methods require a large sample size per subset with $\gamma$ around $0.8$. By contrast, logistic regression can converge with $\gamma$ as low as 0.5, provided enough subsets are used (see the discussion in the next paragraph).

\paragraph{Number of subsets $s$.} Similarly, we see in the error trajectories that large $b$ require only a small number of subsets. In general, similar to \cite{kleiner2014scalable}, we recommend $s \geq 2$ for $\gamma = 0.9$ and $s > 10$ for $\gamma = 0.6$.

\paragraph{Number of bootstrap samples $r$.} The relative error trajectories showed convergence with a fairly small $r = 100$. Similar to \cite{kleiner2014scalable},  Figure \ref{fig:rserr} of the supplementary material shows the relative error as a function of both $r$ and $s$ for logistic regression. We see that the number of bootstrap samples per subset has little effect once we start increasing the number of subsets.
\paragraph{Confidence intervals.} Similar to the traditional bootstrap, confidence intervals can be obtained by using asymptotic normality or by using percentiles \citep{efron1994introduction}. In our simulation, we show the confidence intervals obtained through both the percentile (Figures \ref{fig:zip}, \ref{fig:zip4}, and \ref{fig:zip10}) and asymptotic method (Figures \ref{fig:asympzip}, \ref{fig:asympzip4}, and \ref{fig:asympzip10} of the supplementary material). Our results show that both techniques provide adequate nominal coverage. 
\paragraph{Lack of overlap.} Despite the wide use of IPW estimators, they may lead to extreme weights and erroneous inferences when propensities are close to 0 \citep{kang2007demystifying}. We suggest following standard practice of truncating propensities. Alternatively, we suggest dropping the subset leading to extreme weights and draw a new sample with replacement. 
\paragraph{Covariate balance.} Covariate balance can be computed using standardized mean differences \citep{stuart2010matching} within each bootstrap sample and subset. Practitioners may discard subsets where covariate balance does not reach desirable values and draw a new subset.
\paragraph{Parallel computing.} Causal BLB allows for parallel and distributed implementations using modern computing platforms. Detail of its implementation follows that of the original BLB algorithm of \cite{kleiner2014scalable}.

\section{Simulation}
\label{sec:simulate}


We evaluate the performance of causal BLB using a data-generating process with two covariates and a constant treatment effect, outlined below. We use the presentation guidelines from \cite{morris2019using} to describe the simulation setup.

\begin{align}\label{eq:dgm}
X_1, X_2 &\overset{\textrm{i.i.d.}}{\sim} \textrm{Normal}(0, 1) \nonumber \\ 
\textrm{Pr}(W = 1 | X_1, X_2) &= \dfrac{1}{1 + \exp(- 0.5X_1 - 0.5X_2)} \nonumber \\
\epsilon &\sim \textrm{Normal}(0, 1) \\
Y(0) &= X_1 + X_2 + \epsilon \nonumber \\
Y(1) &= Y(0) + 2W \nonumber
\end{align}

{\bf Data-generating mechanism}: The data-generating mechanism (DGM) is shown in \eqref{eq:dgm}. We draw two independent covariates $(X_{1i}, X_{2i})$ for each subject $i$ and construct the propensity as the inverse-logit of the linear combination of covariates. We then draw an error term $\epsilon$ i.i.d from a standard normal and construct the control and treatment outcomes. The treatment effect is 2, the same for all subjects. The degree of covariate overlap is controlled by the parameter in the inverse-logit equation; $-\frac{1}{2}$ ensures there is some imbalance between the treated and control groups for some of the covariates. 



{\bf Estimand}: Our estimand of interest $\tau$ is the average treatment effect (ATE): $\tau = E(Y(1) - Y(0))$. 

{\bf Methods}: We construct \reps{} replications of the DGM with varying sample size, ranging from 5,000 to 20,000 to examine the histogram of the results and how results vary with increasing sample size.  To illustrate the performance of the causal BLB using several different estimators, we estimate the weights using the propensity score computed by using logistic regression, SVM, and CBPS. 

{\bf Performance Measures}: 
We evaluate (1) timing of the different methods, (2) bias,  and (3) coverage of the bootstrapped confidence intervals. In particular, we want to ensure that the bootstrapped confidence intervals have at least nominal coverage, i.e.,

\[
\textrm{Pr}(\tau \in \textrm{CI}^\ast) \geq 1 - \alpha.
\]

\subsection{Results}

\begin{figure}
    \centering
    \includegraphics[scale=0.90]{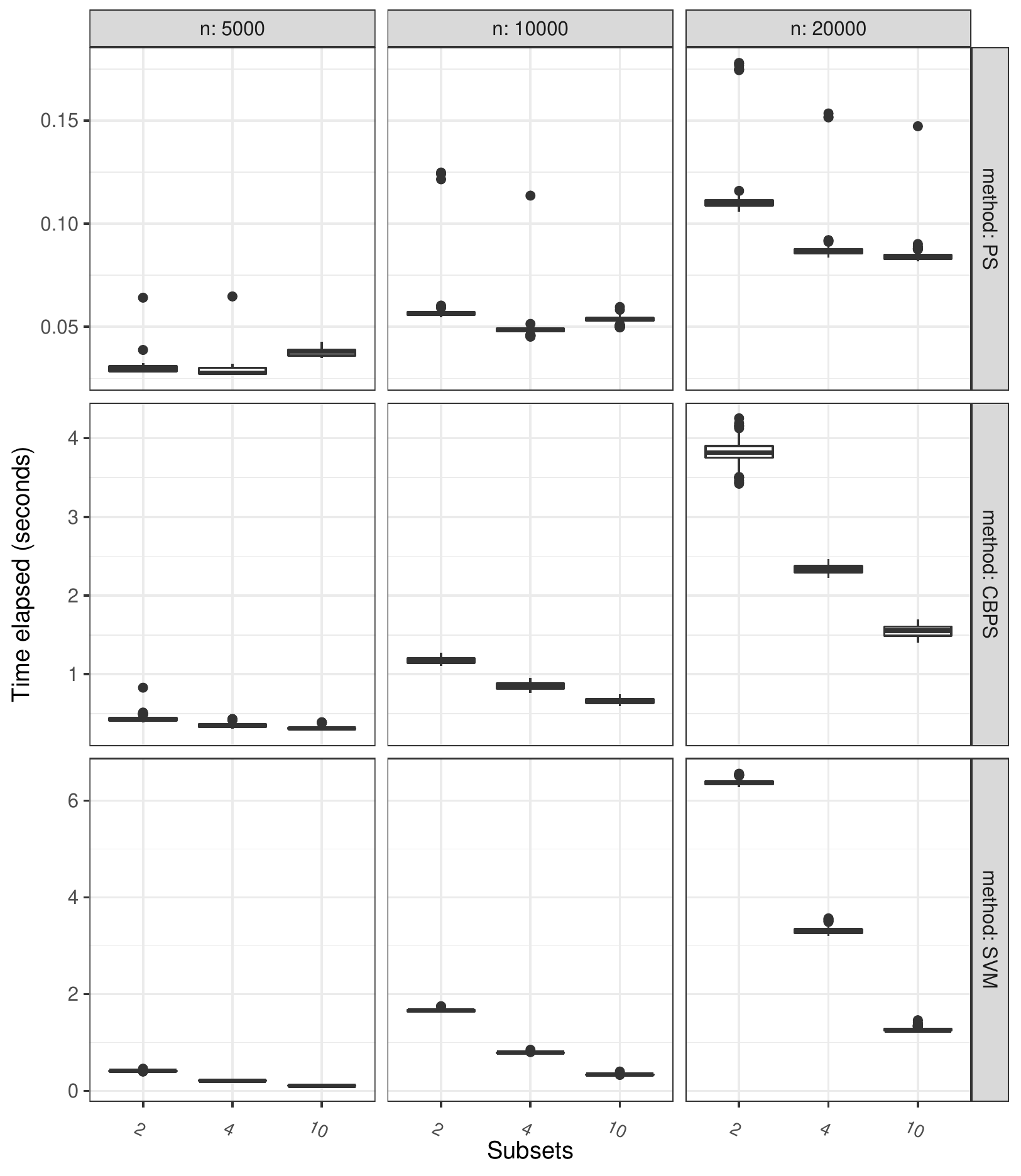}
        \caption{Boxplot of causal BLB times ($B=100$) from \ntimes{} replications.}
    \label{fig:methodtime}
\end{figure}

\paragraph{Timing.} Figure \ref{fig:methodtime} shows the time elapsed for our different propensity score estimation methods by number of subsets. When comparing time taken, we wanted to ensure that the same sample size was used across computations; thus $b$ was set to $n/s$. We see that for computationally intensive methods like CBPS and SVM, there is a distinct time advantage in having more subsets. In addition, we see that there are no real benefits to increasing the number of subsets for methods like logistic regression. We noted this in section \ref{sec:properties} about the causal BLB properties, but the simulation confirms it.


\begin{figure}
    \centering
    \includegraphics[scale=.85]{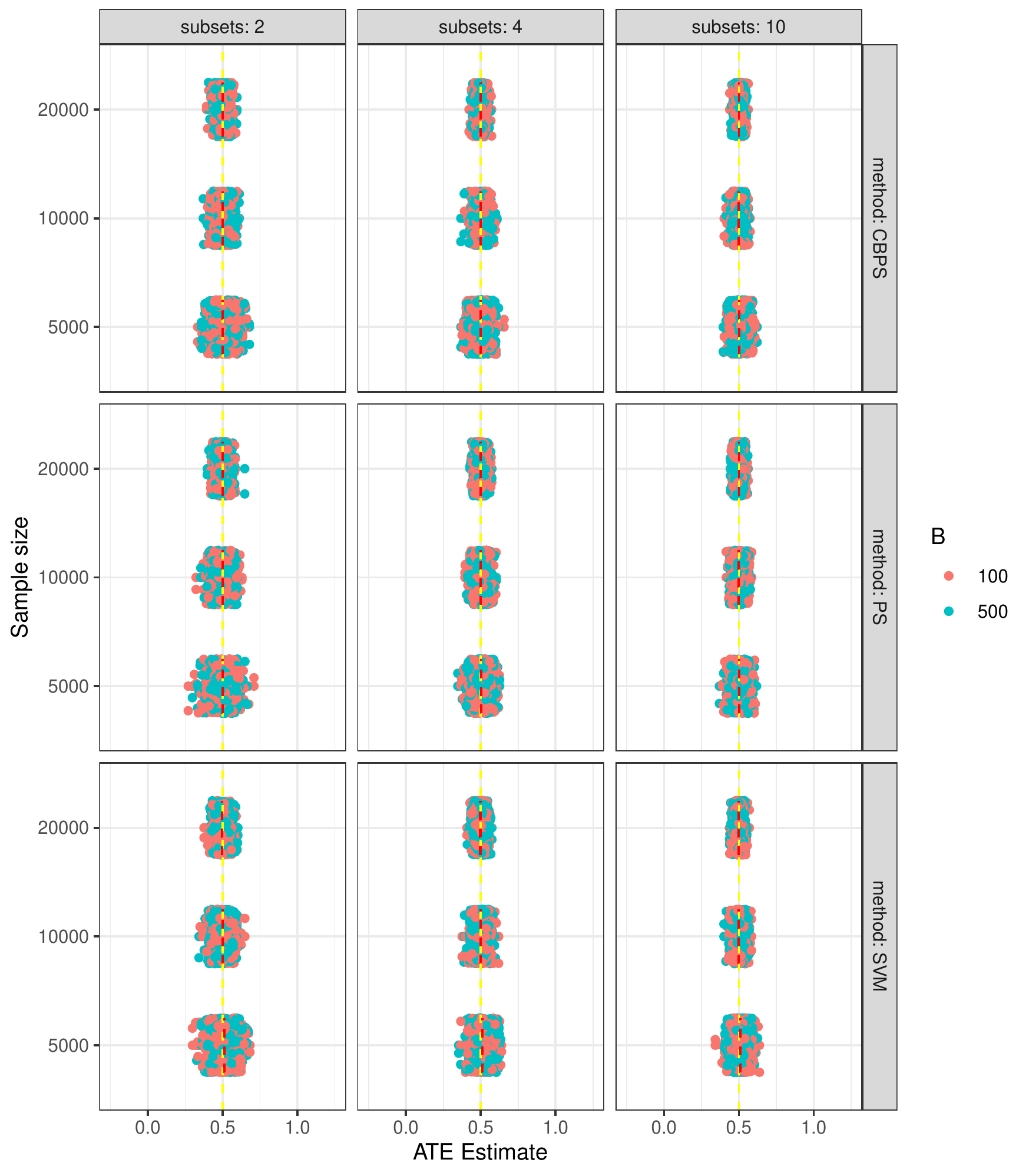}
    \caption{Estimates of ATE from \reps{} replicates (true value is indicated by a dashed line, mean indicated by red line).}
    \label{fig:ates}
\end{figure}

\paragraph{Bias.} Figure \ref{fig:ates} shows the estimates of the ATE from \reps{} replications of the DGM at three different sample sizes, number of subsets, and propensity score weight estimation methods. In this simulation, every computation used $\gamma = 0.8$. We see that all methods produce unbiased estimates.

\begin{figure}
    \centering
    \includegraphics[scale=0.71]{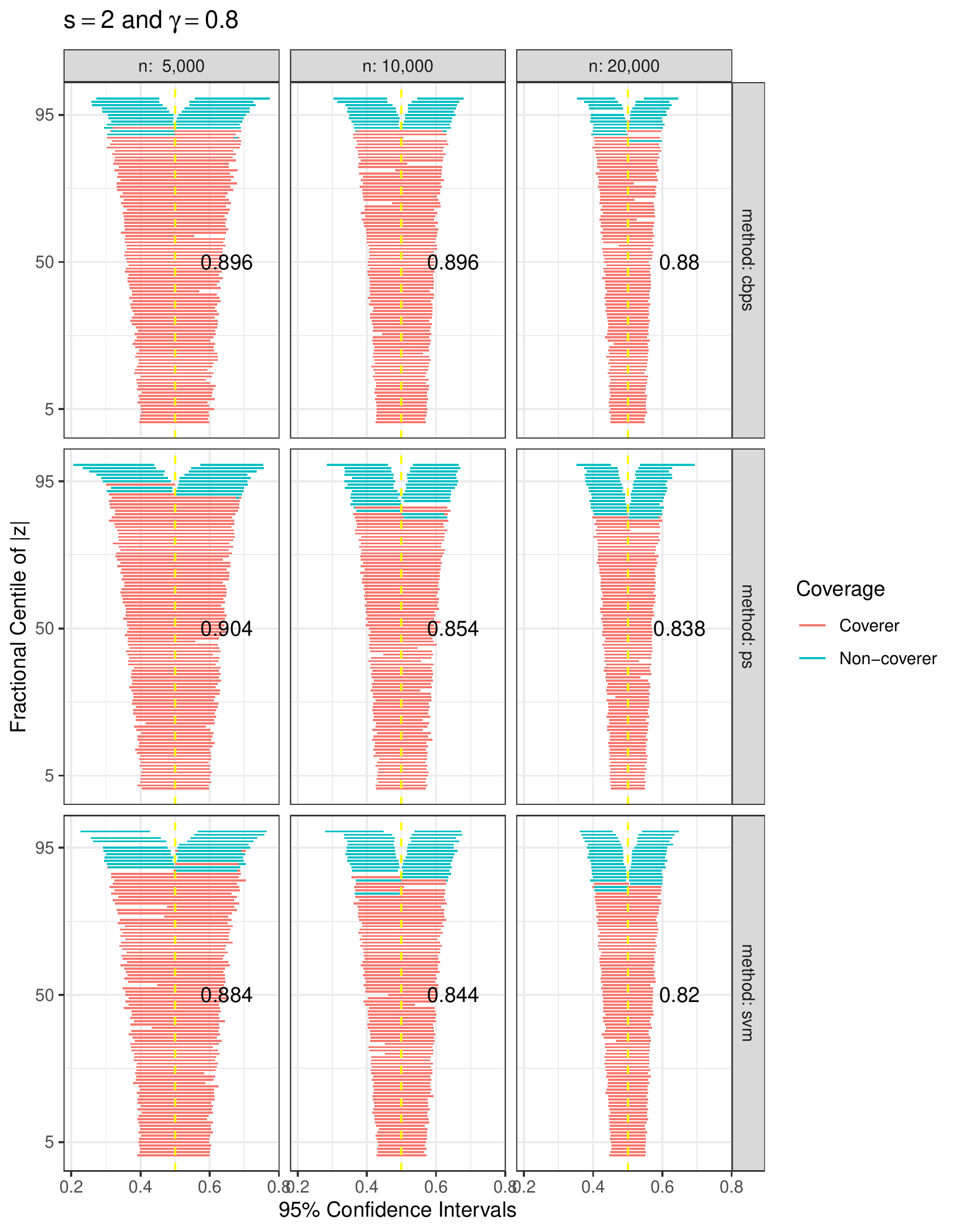}
    \caption{Confidence intervals from \reps{} replications from the causal BLB algorithm, 2 subsets (true value indicate by the yellow line, $B=500$).}
    \label{fig:zip}
\end{figure}

\begin{figure}
    \centering
    \includegraphics[scale=0.71]{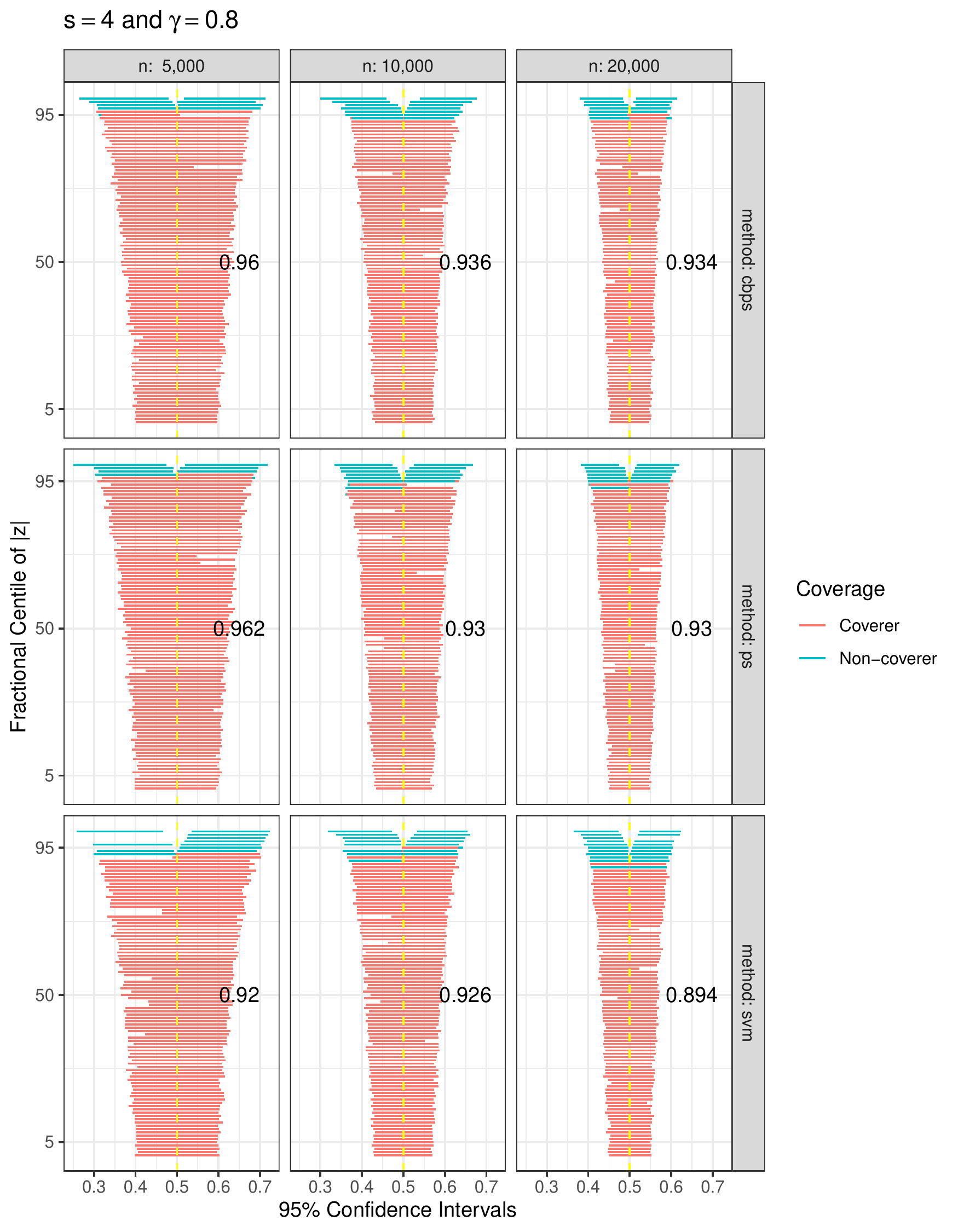}
    \caption{Confidence intervals from \reps{} replications from the causal BLB algorithm, 4 subsets (true value indicate by the yellow line, $B=500$).}
    \label{fig:zip4}
\end{figure}

\begin{figure}
    \centering
    \includegraphics[scale=0.71]{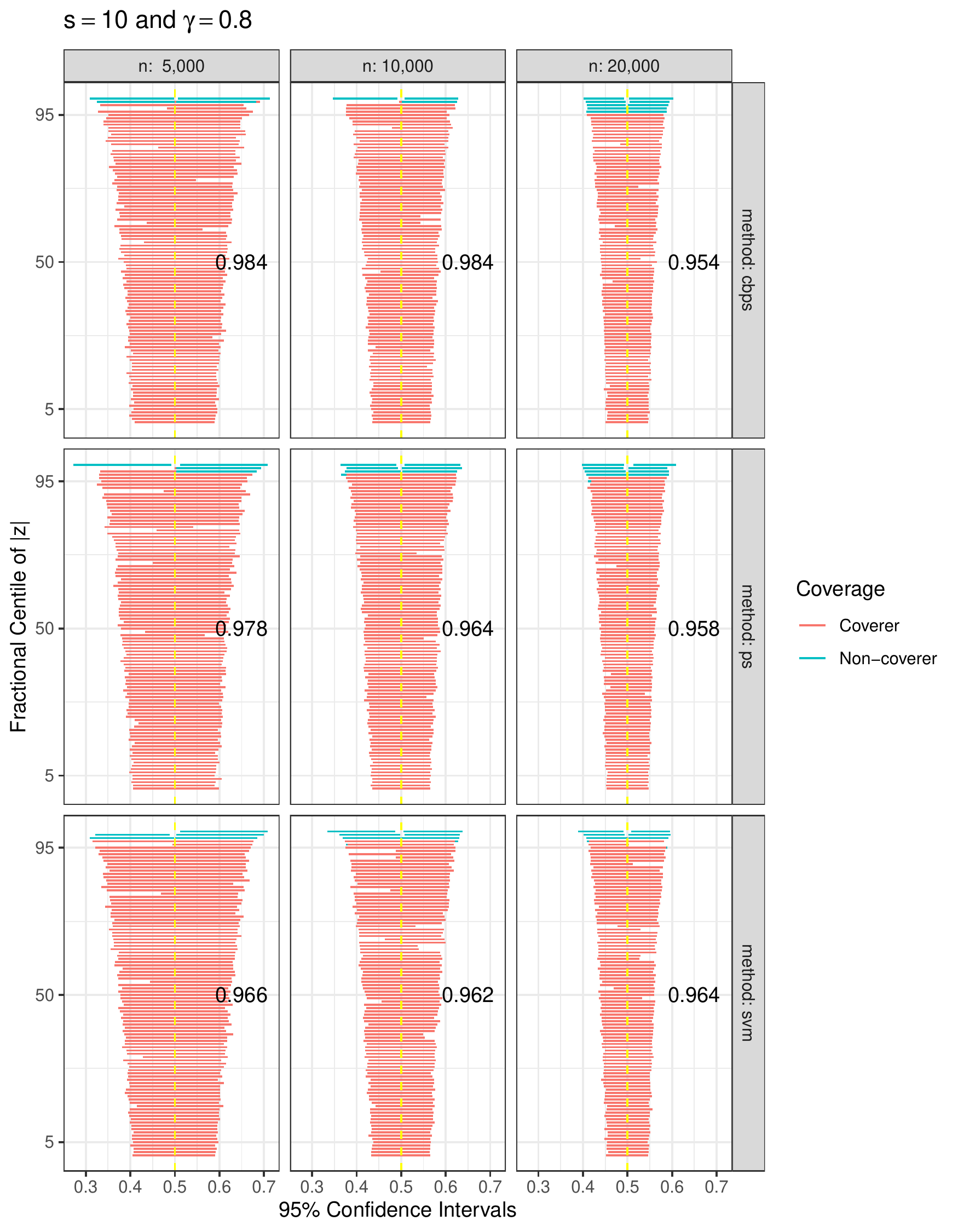}
    \caption{Confidence intervals from \reps{} replications from the causal BLB algorithm, 10 subsets (true value indicate by the yellow line, $B=500$)}
    \label{fig:zip10}
\end{figure}

\paragraph{Coverage} Figures \ref{fig:zip}, \ref{fig:zip4}, and \ref{fig:zip10} shows ``zip plots'' of the percentile confidence intervals for varying subset sizes. These plots display the bootstrap confidence intervals obtained through \reps{} replications. The y-axis shows the fractional centile of $\frac{|\hat{\tau} - 0.5|}{\textrm{se}(\hat{\tau})}$. These figures show that, for $\gamma = 0.8$, $s = 2$ subsets are not sufficient to establish at least the nominal coverage; this problem is more pronounced for the SVM model. As we increase the number of subsets, we see the coverages gets closer to the nominal coverage for $s > 4$. 

Similarly, Figures \ref{fig:asympzip}, \ref{fig:asympzip4}, and \ref{fig:asympzip10} of the supplementary material, show the asymptotic normal \citep{efron1994introduction} confidence intervals, derived for each subset $k$ as:

\[
[\hat{\tau}_k^\textrm{IPW} - z^{1-\alpha}\cdot \textrm{se}(\hat{\tau}_k), \hat{\tau}_k^\textrm{IPW} - z^{\alpha}\cdot \textrm{se}(\hat{\tau}_k)]
\]
\noindent
where $\hat{\tau}_k^\textrm{IPW}$ is the ATE estimator estimated on the entire subset (it is \emph{not} the bootstrap $\hat{\tau}_k$ from algorithm \ref{alg:cblb}) and the standard error is estimated by bootstrap. This is the usual method of calculating asymptotic confidence intervals, see \citet[p. 168]{efron1994introduction}. The standard error here is calculated by taking the standard deviation of ATE estimates over the bootstrap resamples; the confidence interval bounds are then averaged across subsets. We see that the asymptotic results provide similar but slightly more conservative inference.



\section{Application to the Women's Health Initiative}
\label{sec:whi}


To evaluate the causal BLB on a real dataset, we apply it to data from the Women's Health Initiative, a clinical investigation initiated in 1992 to understand the most common causes of mortality and morbidity among postmenopausal women \cite{women1998design}. The initiative consisted of both a randomized trial and observational study. In the randomized trial, 16,608 women were randomly assigned to either daily use of 0.625 mg of conjugated equine estrogen (CEE) and 2.5 mg of medroxyprogesterone acetate (MPA) or a placebo. In the observational study, 53,054 women were enrolled, about a third of whom (17,503) were using combined estrogen-plus-progestin  \cite{prentice2005combined}.

In this section, we evaluate the effect of estrogen plus progestin therapy on average time to coronary heart disease (CHD) among postmenopausal women aged 50-79 years using data from the WHI observational study (September 1993-September 2010). 

\subsection{Study population}\label{sec:studypop}

We considered a target study population of postmenopausal women who in the WHI observational study had reported no use of estrogen therapy, progesterone therapy or a combination of the two during 2-year prior the enrollment in the study. Baseline was defined as first follow-up visit and women were followed from baseline to diagnosis of CHD, loss to follow-up, death, or September 30, 2010, whichever occurred first. Out of the 93,676 women comprising the original WHI observational study 37,080 used any hormone therapy in the 2-year before the enrollment of the study while 30,960 lacked information about the number of days since enrollment, and 1,567 lacked information on time since menopause. The final study population was comprised of 24,069 women.  We considered 34 confounders, listed in section \ref{sec:confound} of the supplementary material. Time since menopause has been recognized as an important factor for the risks and benefits of hormone therapy on CHD \citep{carrasquilla2015association,carrasquilla2017postmenopausal}. We therefore evaluated the impact of estrogen plus progestin therapy on average time to CHD by conducting a stratified analysis on three categories of time since menopause: 0-10 years, 11-20 years and 20+ years.

\subsection{Estimation}

Our goal was to estimate the ATE of hormone therapy on time to CHD using our causal BLB estimator. To obtain propensity scores, we used logistic regression, SVM, and CBPS. For SVM estimation, because the propensity score is likely a complex function of the confounders, we used a radial-basis kernel and a cost parameter of 0.01. Following the simulation results, we run the causal BLB algorithm with $s =2$ and $s = 4$ for each estimation method and each stratified dataset (stratified by time since menopause, more on this in section \ref{sec:studypop}). Rather than choosing the subset size $b$ based on $\gamma$, we chose $b = \frac{n}{s}$ so we can accurately compare the timing of the causal BLB with different subsets. Because each stratified dataset is relatively small (each under 10,000 observations), we omitted the $s = 10$ case. We set the number of bootstrap replications, $r = \frac{500}{s}$. Finally, we calculated the causal BLB estimate as well as the standard error and percentile confidence intervals. The bootstrap standard error is computed by taking the standard deviation of the estimates for each subset and then averaging the results across subsets. For an empirical comparison we also computed a standard IPW estimator using the estimated propensity scores.

\subsection{Results}\label{sec:whiresults}

The results are shown in Table \ref{fig:whiobs}. In this table, 
``Method'' shows the method used to calculate the propensity score. $s$, $b$ and $r$ are the number of subsets, their size, and the number of bootstrap replicates, respectively. ``Full'' indicates the standard IPW estimator estimated on each of the datasets stratified by ``Time since menopause''. $\hat{\tau}$, se($\hat{\tau}$), and ($\textrm{CI}(\tau)$) are the point estimate, standard error and percentile confidence interval computed as described in Algorithm \ref{alg:cblb}. To calculate the computational time, we ran the algorithm \ntimes{} times and then calculated the median time in seconds which is shown in column $\textrm{median}(T)$.\footnote{The point estimate, standard error and CI results are from a single run of the algorithm; the algorithm is only run \ntimes{} times to get an estimate of the time elapsed.} We see that, for each year interval, the results are similar, with a small, negative and statistically non-significant result. The SVM in particular finds a slightly larger negative effect.   Note that contrary to the simulation, CBPS time increases as  $s$ increases and $b$ falls. We had already seen in Figure \ref{fig:methodtime} that the time speed up for causal BLB when using CBPS is fairly small for datasets of size on the order of the stratified WHI datasets. However, the absolute increase in time is a function of the larger number of covariates used in propensity score estimation. The CBPS algorithm with the default number of maximum iterations runs for more iterations in the low $b$, high $s$ case. See section \ref{sec:cbpstime} for further discussion of this.

\begin{table}
\centering
\begin{tabular}[t]{lccccccc}
\toprule
Time since \\ menopause & Method & $s$ & $b$ & Full & $\hat{\tau}$ & se($\hat{\tau}$) ($\textrm{CI}(\tau)$) & $\textrm{median}(T)$\\
\midrule
0-10 & CBPS & 2 & 3708 & 0.00 & 0.00 & 0.05 (-0.1, 0.09) & 6.34\\
0-10 & CBPS & 4 & 1854 & 0.00 & 0.00 & 0.05 (-0.09, 0.09) & 7.77\\
0-10 & PS & 2 & 3708 & -0.01 & -0.03 & 0.04 (-0.11, 0.06) & 0.61\\
0-10 & PS & 4 & 1854 & -0.01 & -0.05 & 0.04 (-0.13, 0.04) & 0.59\\
0-10 & SVM & 2 & 3708 & -0.08 & -0.13 & 0.04 (-0.21, -0.04) & 5.15\\
0-10 & SVM & 4 & 1854 & -0.08 & -0.08 & 0.04 (-0.17, 0) & 2.80\\
$<$10-20 & CBPS & 2 & 4768 & 0.00 & 0.00 & 0.06 (-0.11, 0.1) & 8.73\\
$<$10-20 & CBPS & 4 & 2384 & 0.00 & 0.00 & 0.05 (-0.11, 0.09) & 11.17\\
$<$10-20 & PS & 2 & 4768 & -0.05 & -0.07 & 0.05 (-0.17, 0.04) & 0.76\\
$<$10-20 & PS & 4 & 2384 & -0.05 & -0.08 & 0.05 (-0.16, 0.02) & 0.75\\
$<$10-20 & SVM & 2 & 4768 & -0.13 & -0.19 & 0.05 (-0.28, -0.09) & 6.42\\
$<$10-20 & SVM & 4 & 2384 & -0.13 & -0.19 & 0.05 (-0.28, -0.1) & 3.52\\
20+ & CBPS & 2 & 3434 & 0.00 & 0.00 & 0.08 (-0.16, 0.14) & 8.44\\
20+ & CBPS & 4 & 1717 & 0.00 & -0.01 & 0.09 (-0.19, 0.15) & 14.46\\
20+ & PS & 2 & 3434 & -0.06 & -0.03 & 0.1 (-0.22, 0.16) & 0.61\\
20+ & PS & 4 & 1717 & -0.06 & -0.05 & 0.09 (-0.21, 0.12) & 0.62\\
20+ & SVM & 2 & 3434 & -0.15 & -0.05 & 0.09 (-0.23, 0.12) & 2.37\\
20+ & SVM & 4 & 1717 & -0.15 & -0.27 & 0.1 (-0.45, -0.08) & 1.34\\
\bottomrule
\end{tabular}
    \caption{\footnotesize{Application of causal BLB to estimate the ATE of hormone therapy on time to coronary heart disease using data from the WHI observational study. ``Method'' for the propensity score. $s$, $b$ and $r$ are the number of subsets, their size, and the number of bootstrap replicates ($r = \frac{500}{s}$)}. ``Full'' indicates the IPW estimate stratified by ``Time since menopause''. $\hat{\tau}$, se($\hat{\tau}$), and ($\textrm{CI}(\tau)$) are the point estimate, standard error and percentile confidence interval computed as described in Algorithm \ref{alg:cblb}. Median time in seconds.}
    \label{fig:whiobs}
\end{table}




\section{Conclusions}
\label{sec:conc}


In this paper, we introduced causal BLB, a novel bootstrap algorithm that improves the computational time of the traditional bootstrap while consistently estimating causal effects from large data. 
We demonstrated that the proposed estimator obtained is similar to a variance stabilized IPW estimator. We also showed that the technique offers favorable computational advantages for complex machine learning methods like SVM. The main limitation of our method is that we must correctly estimate the propensity score model. As a result, further work will include extending causal BLB to doubly robust methods. In addition, based on our promising finite-sample simulation results, future directions include demonstrating that causal BLB leads to asymptotically correct confidence intervals.  

\bibliography{biblio}
\newpage

\appendix

\section{Consistency of bootstrap estimator}\label{sec:consistent}



For each subset $k$, order the data such that the all control units are followed by all treatment units. The normalized IPW estimator for the ATE in subset $k$ is:

\begin{align}\label{eq:ipw}
\hat{\tau}_k &= \sum_{i = 1}^b \left(\dfrac{W_i Y_i w^1_{i,k}(X_i)}{\sum_{i=1}^b W_i w^1_{i,k}(X_i)} - \dfrac{(1-W_i) Y_i w^0_{i,k}(X_i)}{\sum_{i=1}^b (1-W_i) w^0_{i,k}(X_i)}\right) \\
&= \sum_{i = 1}^b \left(W_i Y_i w^{1'}_{i,k}(X_i) - (1-W_i) Y_i w^{0'}_{i,k}(X_i)\right) 
\end{align}

Suppose in subset $k$ there are $b_k^1$ treated units and $b_k^0 \equiv b - b_k^1$ controls. Now, for each replicate $j$, consider the following multinomial draws: 
\begin{align*}
\left(M^1_{1,j,k}, \ldots, M^1_{b_k^1, j, k}\right) &\sim \textrm{Multinomial}(n_1; w^{1'}_{1,k}(X_1), \ldots, w^{1'}_{b_k^1,k}(X_{b_k^1})) \\
\left(M^0_{1,j,k}, \ldots, M^0_{b_k^0, j, k}\right) &\sim \textrm{Multinomial}(n_0; w^{0'}_{1,k}(X_1), \ldots, w^{0'}_{b_k^0,k}(X_{b_k^0}))
\end{align*}

Now construct the $\hat{\tau}_{j,k}$

\begin{align*}
\hat{\tau}_{j,k} &= \sum_{i=1}^{b}\left(\dfrac{-1}{n_0}\right)^{W_{i,k} - 1}\left(\dfrac{1}{n_1}\right)^{W_{i,k}}\Tilde{M}_{i,j,k}Y_{i,k} 
\end{align*}


Conditional on the observed data in the subset, the expectation of the multinomial draws vectors is:

\[
E(\Tilde{M}_{i,j,k}) = \begin{cases}
    n_1w^{1'}_{i,k} & \textrm{if } W_i = 1 \\
    n_0w^{0'}_{i,k} & \textrm{if } W_i = 0
\end{cases}
\]

Thus, for one term of the above summation,

\begin{align*}
\textrm{E}_{\mathbf{M}}\left(\left(\dfrac{-1}{n_0}\right)^{W_{i,k} - 1}\left(\dfrac{1}{n_1}\right)^{W_{i,k}}\Tilde{M}_{i,j,k}Y_{i,k}
\right) &= \dfrac{W_i Y_i n_1w^{1'}_{i,k}}{n_1} - \dfrac{(1 - W_i) Y_i n_0w^{0'}_{i,k}}{n_0} \\
&= W_i Y_i w^{1'}_{i,k} - (1 - W_i) Y_i w^{0'}_{i,k}
\end{align*}


Thus, by the weak law of large numbers, over $r$ weighted bootstrap resamples (corresponding to $r$ treatment and control multinomial draws)

\[
\dfrac{1}{r} \sum_{i=1}^r \hat{\tau}_{i,k}\overset{p}{\to} \hat{\tau}_k
\]



Because each subset is randomly chosen subset of the full dataset, under the usual causal inference assumptions and assuming the propensity score is correctly specified, $\hat{\tau}_k \inprob \tau$.

\section{Additional Parameter Considerations}
\subsection{Relative Error Simulation}
\label{sec:hyperparam}

To determine practical guidelines, we conducted simulations similar to \citet{kleiner2014scalable}. We use the same DGM as described in section \ref{sec:simulate}.  First, we created \truereps{} independent replications of our simulation dataset with $n=20000$. On each dataset, we estimated our IPW variance-stabilized ATE estimator and then quantiled the results to get the ``true'' confidence interval for the variance-stabilized IPW estimator. Then, on an independent dataset, we iteratively ran our causal BLB algorithm with the number of bootstrap replicates per subset set at $r = 100$, updated our ATE estimate and recorded the time it took to process each subset. With each updated estimate, we calculated the relative error as follows: Let $\xi_\textrm{lo}$ and $\xi_\textrm{up}$ be the lower and upper bounds respectively of the true confidence interval and let $c_\textrm{lo}^s$ and $c_\textrm{up}^s$ be the lower and upper bounds of the BLB confidence interval after processing the $s^\textrm{th}$ subset. Then the relative error metric after processing subset $s$ is:

\[
\textrm{Err}^s = \dfrac{\frac{\left|c_\textrm{lo}^s- \xi_\textrm{lo}\right|}{\xi_\textrm{lo}} + \frac{\left|c_\textrm{up}^s- \xi_\textrm{up}\right|}{\xi_\textrm{up}}}{2}
\]

This was repeated \datareps{} times on independent data realizations, with the estimates and times at each iteration averaged together. These simulations produced a trajectory of relative error as a function of time. When these are plotted, we can visually inspect the time to ``convergence'' for varying values of $\gamma$.

\begin{figure}
    \centering
    \includegraphics{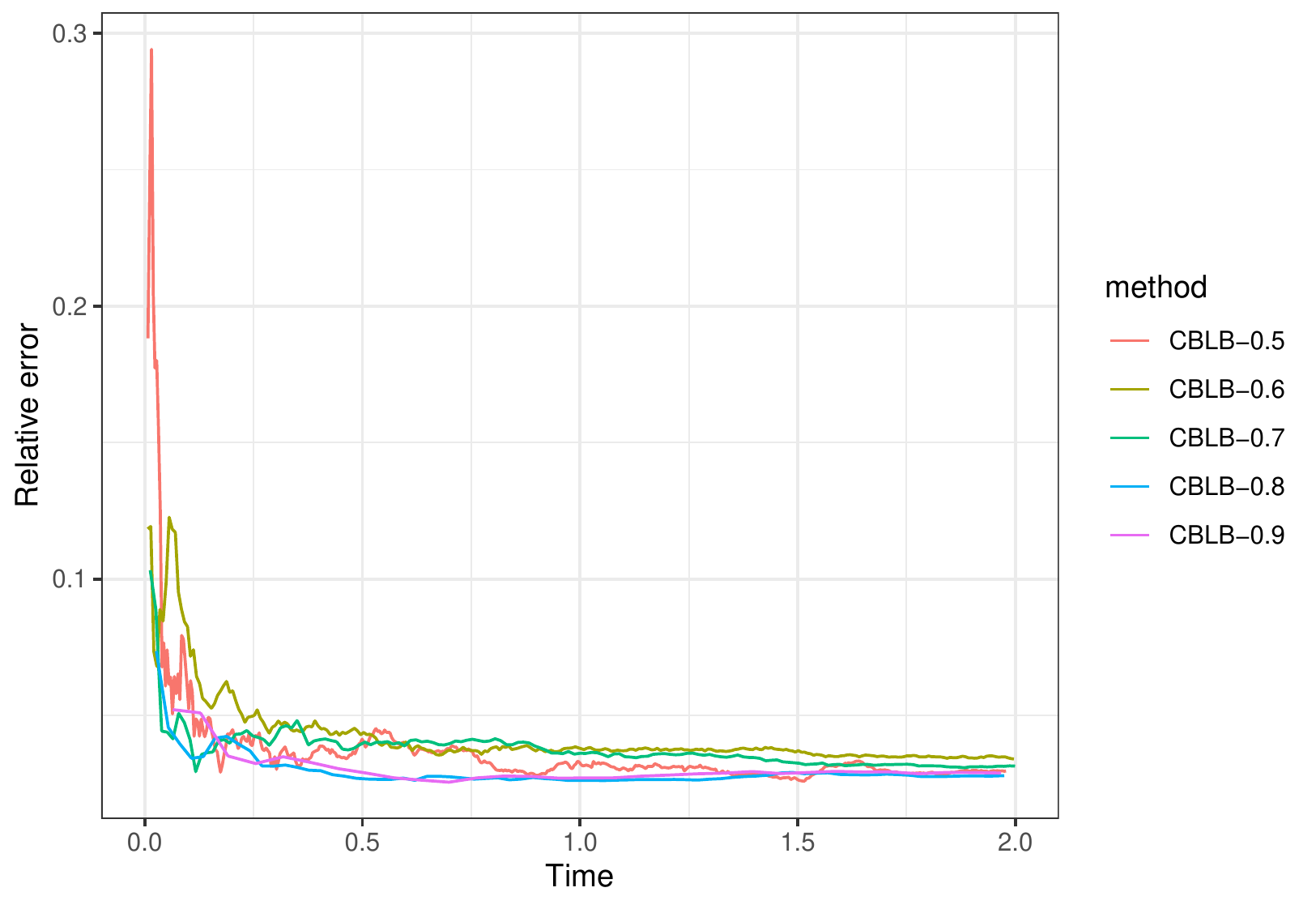}
    \caption{Relative error trajectories vs. processing time for $n = 20000$, propensity score weights estimated by logistic regression.}
    \label{fig:logiterr}
\end{figure}

\begin{figure}
    \centering
    \includegraphics{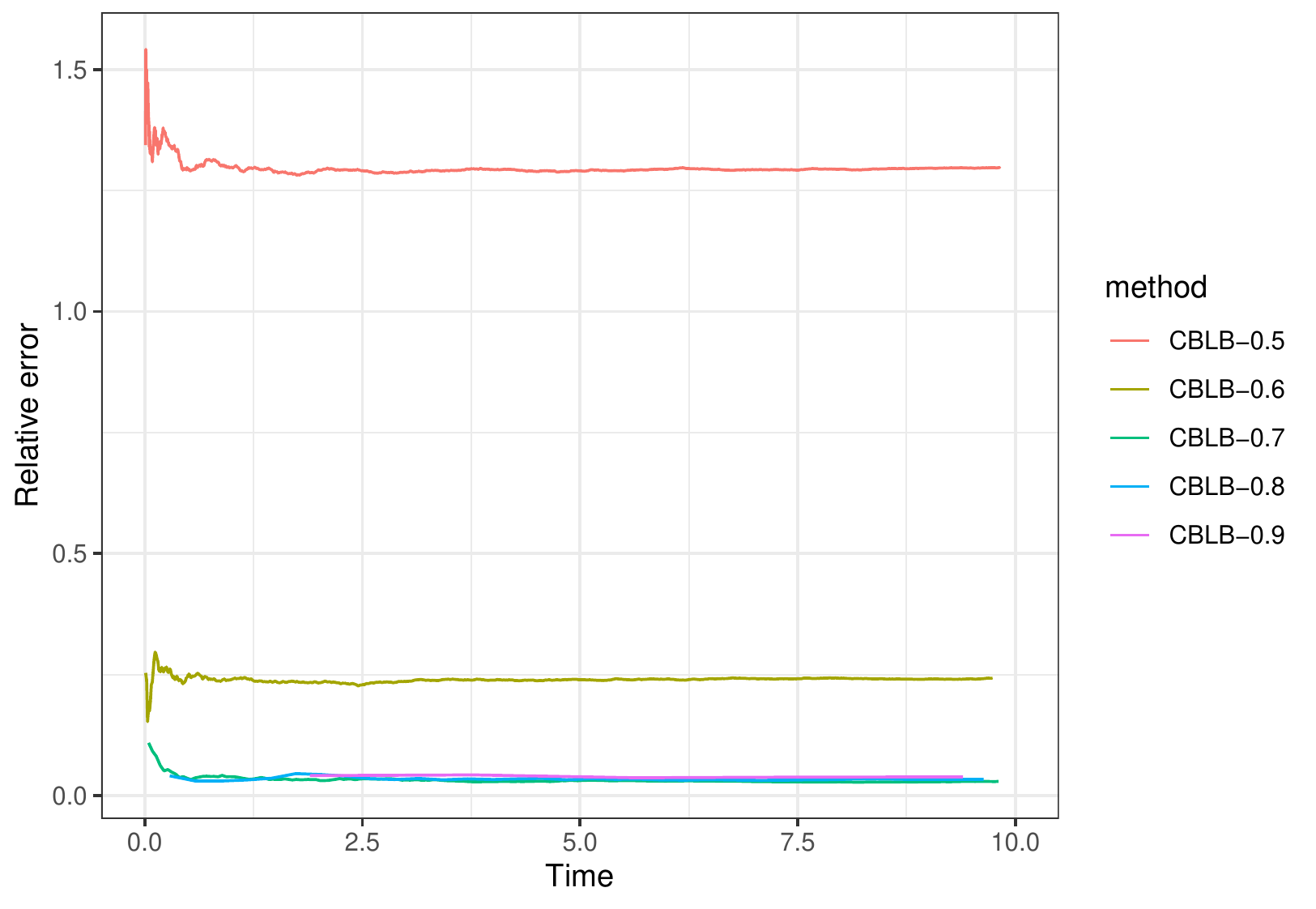}
    \caption{Relative error trajectories vs. processing time for $n = 20000$, propensity score weights estimated by support vector machine.}
    \label{fig:svmerr}
\end{figure}

\begin{figure}
    \centering
    \includegraphics[scale=0.85]{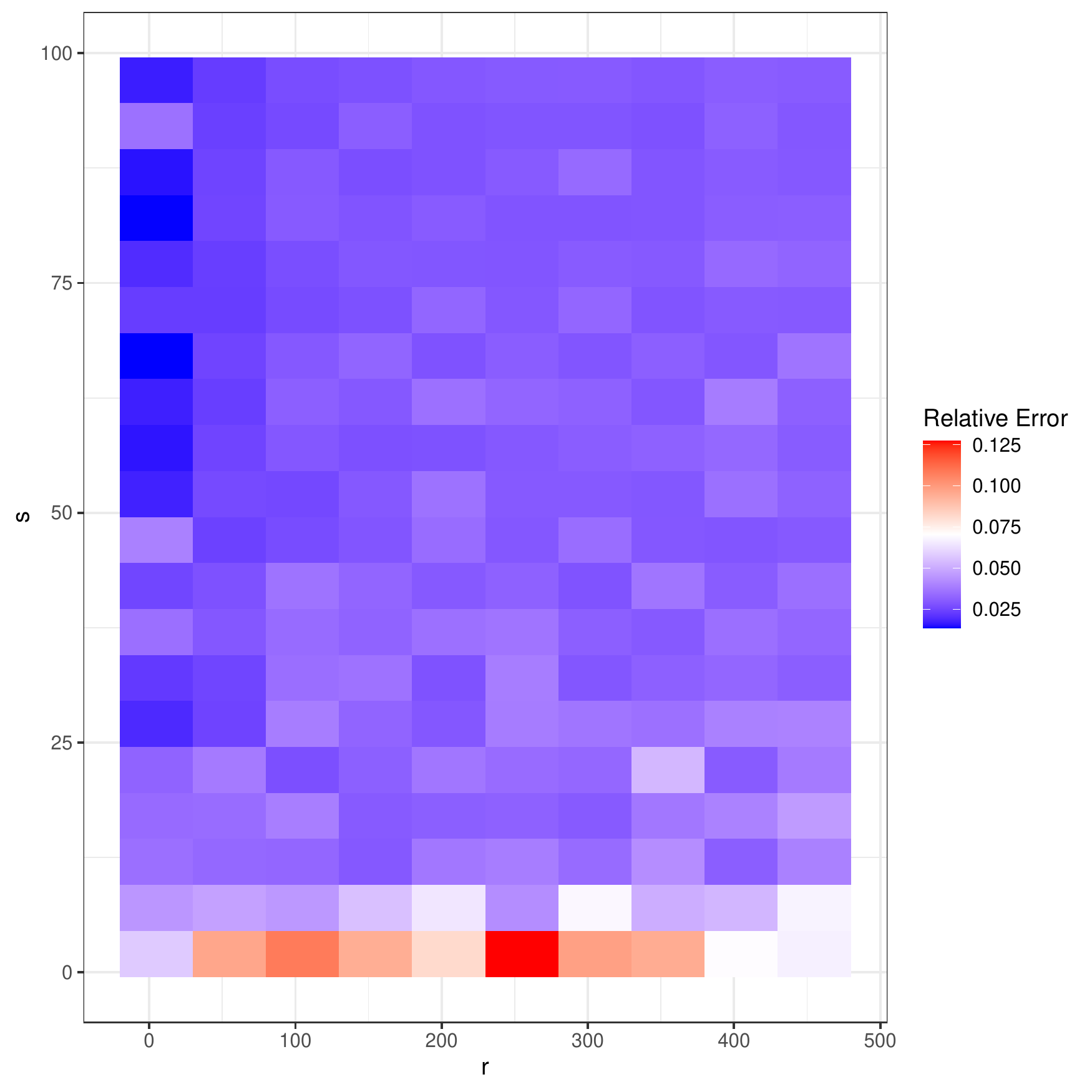}
    \caption{Relative error achieved for different values of $r$ and $s$, $n = 20000$, propensity score weights estimated by logistic regression.}
    \label{fig:rserr}
\end{figure}

Figure \ref{fig:logiterr} shows the relative error trajectory for $\gamma \in \{0.5, 0.6, 0.7, 0.8, 0.9\}$ where the propensity score weights are calculated using logistic regression. Similarly, Figure \ref{fig:svmerr} shows the same trajectory, except with the propensity scores estimated using SVM with a linear kernel and cost parameter equal to 0.01. Figure \ref{fig:rserr} shows the relative error achieved by the causal BLB for different values of $r$ and $s$.

\section{Asymptotic Confidence Intervals}

In this section, we include the asympotic confidence intervals.
\begin{figure}
    \centering
    \includegraphics[scale=0.71]{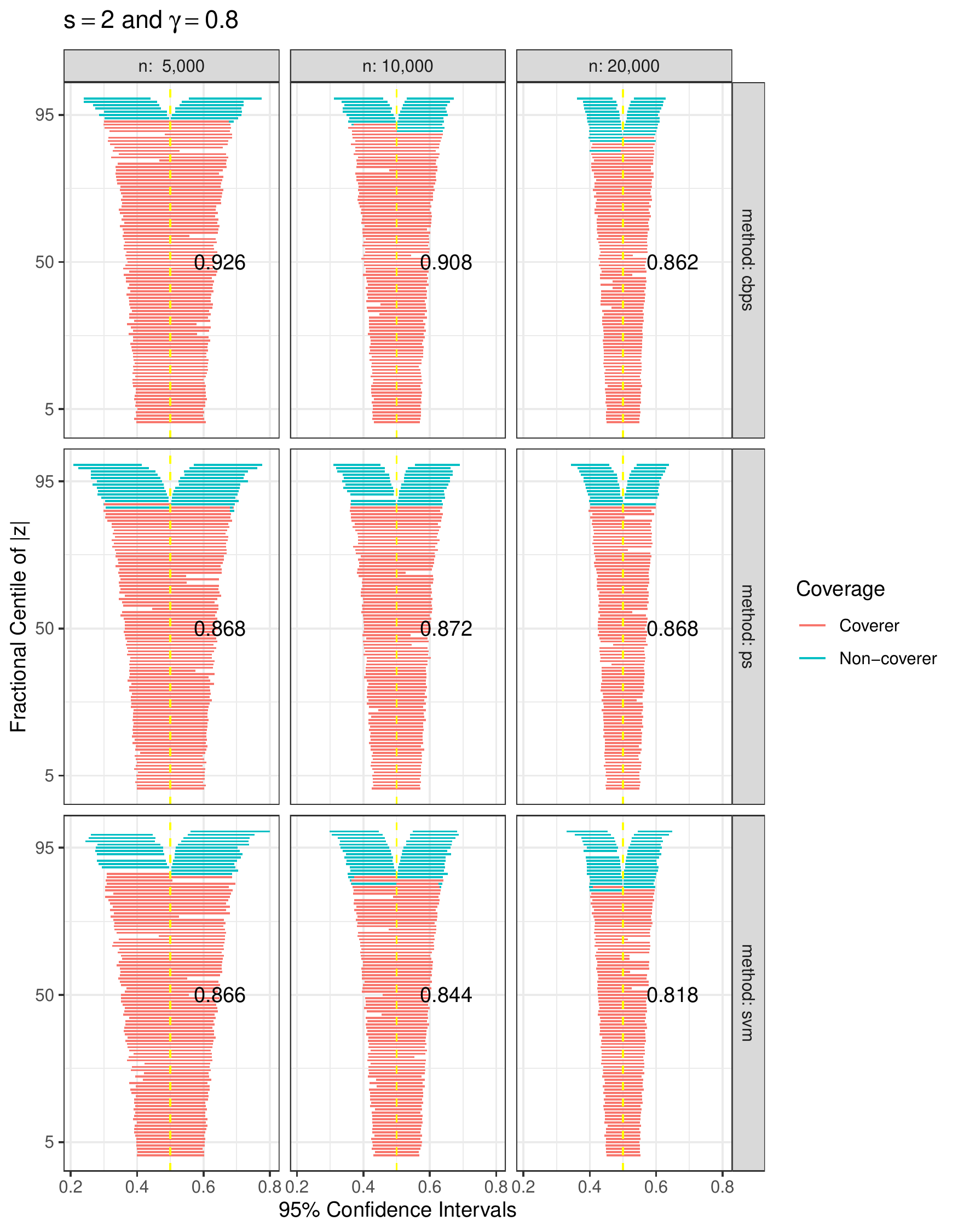}
    \caption{Asymptotic confidence intervals from \reps{} replications from the causal BLB algorithm, 2 subsets (true value indicate by the yellow line, $B=500$).}
    \label{fig:asympzip}
\end{figure}

\begin{figure}
    \centering
    \includegraphics[scale=0.71]{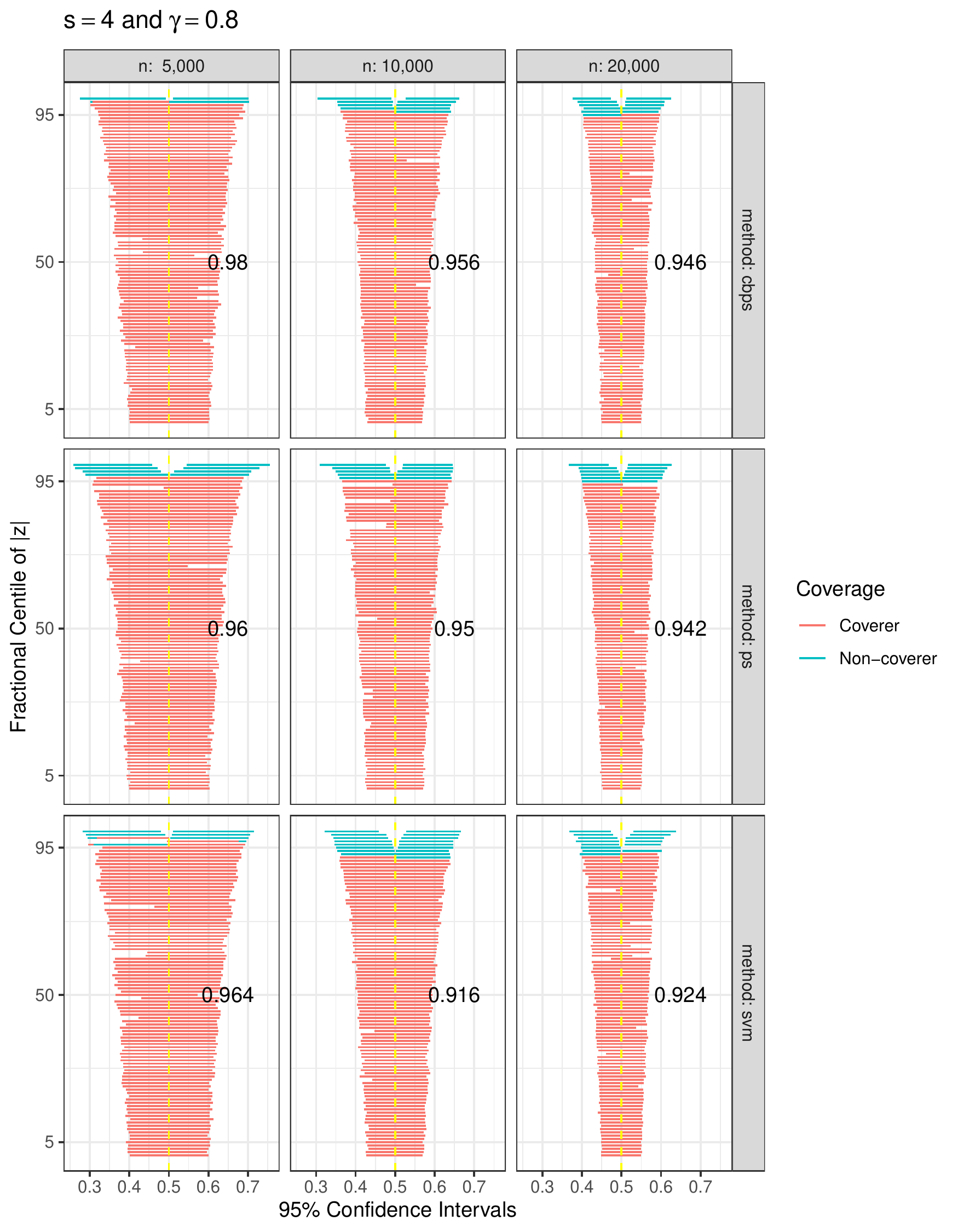}
    \caption{Asymptotic confidence intervals from \reps{} replications from the causal BLB algorithm, 4 subsets (true value indicated by the yellow line, $B=500$).}
    \label{fig:asympzip4}
\end{figure}

\begin{figure}
    \centering
    \includegraphics[scale=0.71]{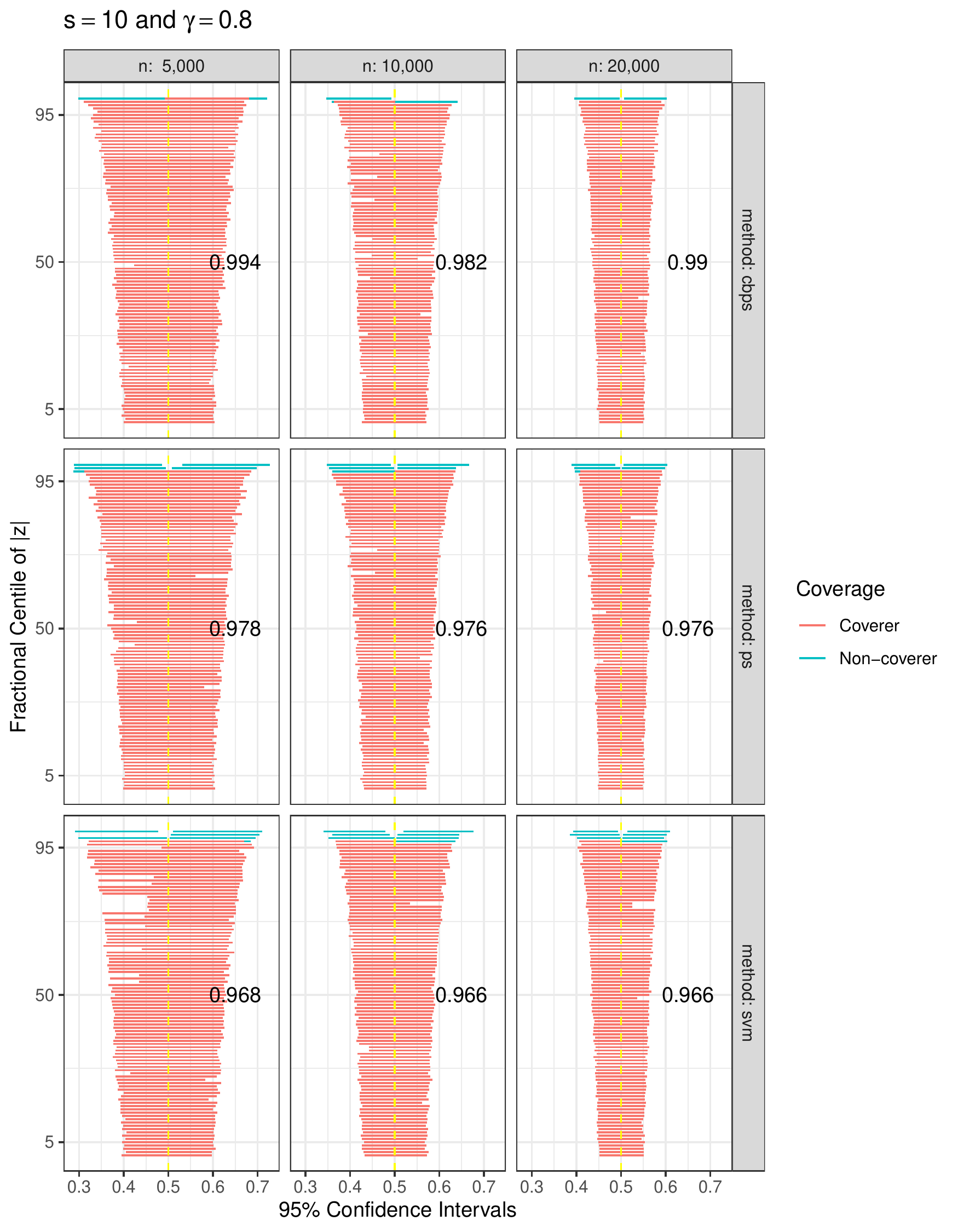}
    \caption{Asymptotic confidence intervals from \reps{} replications from the causal BLB algorithm, 10 subsets (true value indicate by the yellow line, $B=500$).}
    \label{fig:asympzip10}
\end{figure}

\section{Confounders}
\label{sec:confound}

The list of confounders for the WHI observational data is as follows:

\begin{itemize}
    \item multivitamin without minerals use (yes, no)
    \item multivitamin with minerals use (yes, no), 
    \item ethnicity (White, Black, Hispanic, Native American, Asian/Pacific Islander, Unknown), \item number of pregnancies (7 categories)
    \item bilateral oophorectomy (yes, no),
    \item age at menopause (numeric), \item breast cancer ever (yes, no)
    \item colon cancer ever  (yes, no) \item endometrial cancer ever  (yes, no)
    \item skin cancer ever  (yes, no) 
    \item melanoma cancer ever (yes, no)
    \item other cancer past 10 years  (yes, no)
    \item deep vein thrombosis ever (yes, no)
    \item stroke ever (yes, no), myocardial infarction ever (yes, no)
    \item diabetes ever (yes, no)
    \item high cholesterol requiring pills ever (yes, no)
    \item osteoporosis ever  (yes, no) 
    \item cardiovascular disease ever  (yes, no)
    \item coronary artery bypass graft  (yes, no)
    \item atrial fibrillation ever  (yes, no)
    \item aortic aneurysm ever  (yes, no)
    \item angina  (yes, no), hip fracture age 55 or older  (yes, no)
    \item smoked at least 100 cigarettes ever (yes, no)
    \item alcohol intake (non drinker, past drinker,  less than 1 drink per month, less than 1 drink per week, 1 to 7 drinks per week, 7+ drinks per week)
    \item fruits med serv/day (numeric)
    \item vegetables med serv/day (numeric)
    \item dietary energy (kcal)
    \item systolic blood pressure (numeric)
    \item diastolic blood pressure (numeric)
    \item body mass index (numeric)
    \item education (11 categories)
    \item income (10 categories).
\end{itemize}

\section{CBPS Time}\label{sec:cbpstime}

\begin{figure}
    \centering
    \includegraphics{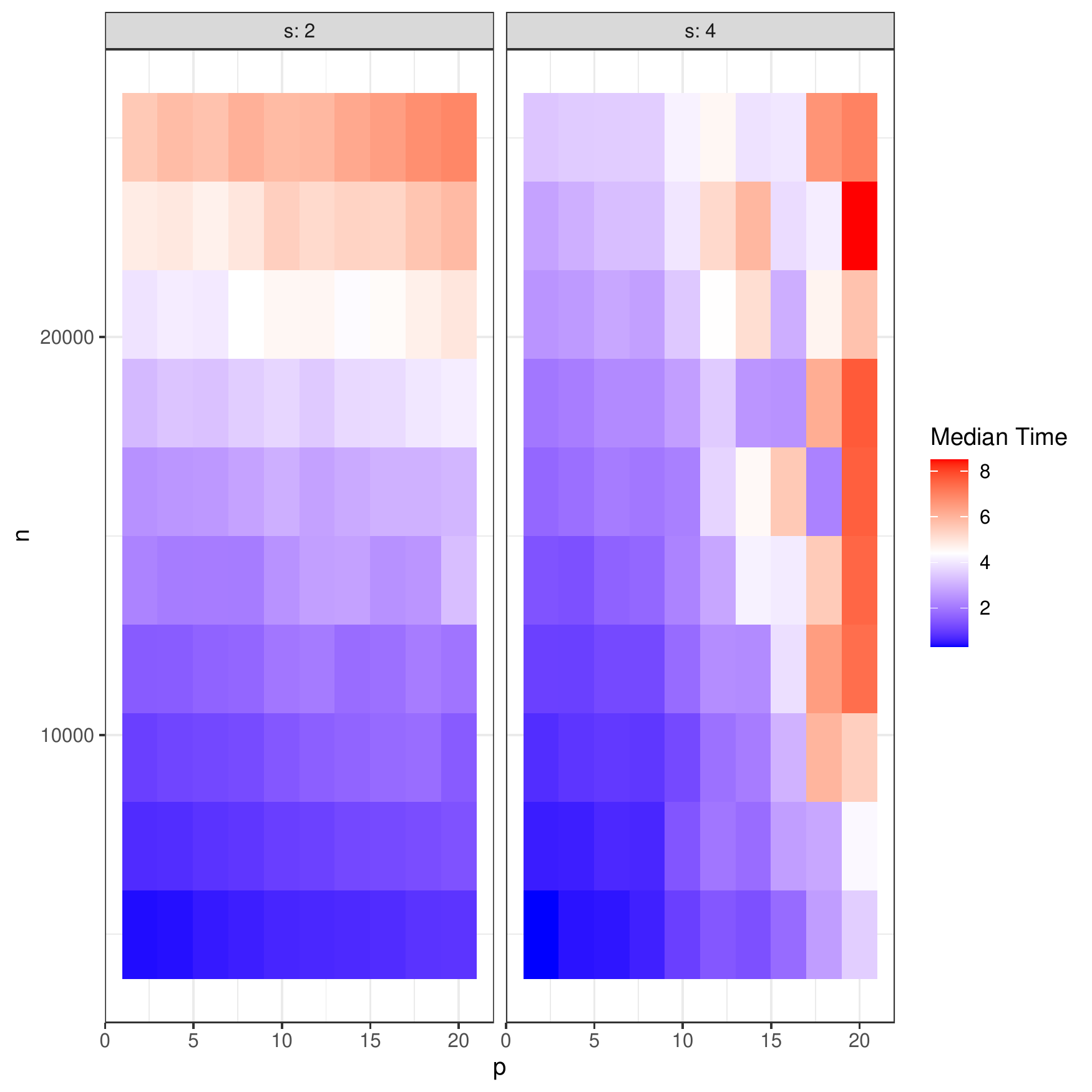}
    \caption{CBPS median time as a function of $n$ and $p$.}
    \label{fig:cbpstimechart}
\end{figure}

To examine the performance of CBPS in the presence of multiple confounders, we constructed the following simulation. We created a simple model with a binary outcome $y$ and probability of success determined by a linear combination of $p$ predictors. We treated $y$ as a treatment indicator and the predictors $\mathbf{X}$ as confounders. To estimate how causal BLB performs for different $n$-$p$ combinations, we construct two datasets one of size $n/2$ and one of size $n/4$. We run the CBPS algorithm 2 times on the first and 4 times on the second to mimic the action of causal BLB on subsets of the data (because with $s$ subsets, the CBPS runs $s$ times). Each $n$-$p$ combination is run 10 times and the median time is calculated.  The results are shown in Figure \ref{fig:cbpstimechart}. We see that, for a small number of predictors, causal BLB  with $s = 4$ can improve on $s=2$, but requires more data as the number of predictors $p$ increases.

\end{document}